\documentclass[openacc]{rsproca_new}
\definecolor{green}{RGB}{0,100,0}
\usepackage{graphicx}
\newcommand\arcsec{\mbox{$^{\prime\prime}$}}%
\newcommand{\rev}[1]{#1}

\titlehead{Research}

\begin{document}

\title{Signs of Sulphur fractionation under high magnetic field strength}

\author{
Dominik Orlovskij$^{1}$, Andy S.~H.~To$^{2}$ and David M.~Long$^{1}$
}

\address{$^{1}$Centre for Astrophysics \& Relativity, School of Physical Sciences, Dublin City University, Glasnevin Campus, Dublin, D09 V209, Ireland\\
$^{2}$European Space Agency (ESA), European Space Research and Technology Centre (ESTEC), Keplerlaan 1, 2201 AZ Noordwijk, the Netherlands\\
}

\subject{Solar physics, Astrophysics, Elemental fractionation}


\corres{Dominik Orlovskij\\
\email{dominik.orlovskij2@mail.dcu.ie}}

\begin{abstract}
Sulphur, with a first ionisation potential (FIP) of 10.36~eV, lies at the boundary between low- and high-FIP elements, making it particularly sensitive to fractionation processes in the solar atmosphere. Sulphur exhibits variable behaviour across solar environments, with coronal remote sensing studies often observing it as a high-FIP element while in-situ measurements sometimes detect low-FIP-like enhancement. Sulphur also exhibits variable behaviour during flares and magnetic restructuring. To understand sulphur's variations, we quantify how sulphur's FIP bias depends on potential field source surface (PFSS)-derived loop properties. We analyse nine Hinode/EUV Imaging Spectrometer (EIS) raster observations using four diagnostic line pairs (Si~\textsc{x} 258.37~\AA\ / S~\textsc{x} 264.23~\AA, S~\textsc{xi} 188.68~\AA\ / Ar~\textsc{xi} 188.81~\AA, Ca~\textsc{xiv} 193.87~\AA\ / Ar~\textsc{xiv} 194.40~\AA, and Fe~\textsc{xvi} 262.98~\AA\ / S~\textsc{xiii} 256.69~\AA), with FIP biases derived using differential emission measures (DEM) calculated via regularised inversion. Our results show that abundances of low-FIP elements, including sulphur, decrease above $\sim$150~G relative to the high-FIP element Ar, while showing no dependence on loop length. This provides evidence that FIP fractionation is modulated by mean magnetic field strength of coronal loops.
\end{abstract}

\rsbreak

\section{Introduction}

The elemental composition of solar plasma is an indicator of the physical processes (such as transportation of energy and mass) that occur between the different layers of the solar atmosphere. The plasma composition of the photosphere is relatively well established~\cite{2009ARA&A..47..481A}. In contrast, the composition of the solar corona often shows elemental abundances that differ from photospheric values~\cite{Scott2015A&A...573A..25S}, a phenomenon known as the first ionisation potential (FIP) effect~\cite{Pottasch1963Apr, Baker2013Nov,Brooks2015}. In the corona, high-FIP elements (FIP $\geq$ 10~eV; e.g., O, Ar, and Ne) typically remain near their photospheric abundances, whereas the abundance of low-FIP elements (FIP $<$ 10~eV; e.g., Ca, Mg, Fe, Si) are typically enhanced in active regions (ARs). This enhancement is quantified by the FIP bias. The absolute FIP bias of element \textit{i} is defined as the ratio of an element's coronal abundance to its photospheric abundance:
\begin{align}\label{1}
\mathrm{FIP\ bias}_i = \frac{A_{i,\ \mathrm{corona}}}{A_{i,\ \mathrm{photosphere}}}.
\end{align}

\rev{However, extreme-ultraviolet (EUV) diagnostics cannot directly measure the photospheric abundances of an element ($\mathrm{A_{i, photosphere}}$). Therefore, FIP bias values derived from EUV observations (e.g., Hinode/Extreme-ultraviolet Imaging Spectrometer, EIS) are relative rather than absolute.} \rev{In this case, relative FIP bias assumes that the high-FIP element in the solar atmosphere remains at its photospheric abundance. Under this assumption,}
\begin{align}\label{2}
\mathrm{Relative\ FIP\ bias}_{i, j} = \frac{A_{i,\ \mathrm{corona}}}{A_{j,\ \mathrm{corona}}},
\end{align} 
\rev{where the coronal abundance of a low-FIP element~\textit{i} is measured against a high-FIP element~\textit{j}: Relative FIP bias values are typically 3--4 in AR cores, indicating that the abundance of low-FIP elements are enhanced by factors of 3--4 relative to the high-FIP element.} However, this represents a simplified picture, as significant spatial and temporal variations exist across different solar structures. In some localised regions such as light bridges or flare sites, an inverse-FIP (IFIP) effect has been observed (i.e., low-FIP elements become depleted relative to high-FIP), with EIS measurements showing Ca/Ar FIP bias values dropping to as low as 0.4--0.5~\cite{Doschek2015Jul, Doschek2016Jun, Brooks2018ApJ...863..140B, Brooks2022ApJ...930L..10B,Baker2024ApJ...970...39B}. 



\rev{Currently, the theoretical framework that best explains the observed FIP effect is the ponderomotive force model~\cite{2004ApJ...614.1063L,2009ApJ...695..954L,2015LRSP...12....2L}.} In this framework, Alfv\'{e}n waves in the corona propagate downwards along closed magnetic field lines and encounter the steep density gradient of the transition region before the upper chromosphere. These gradients cause the waves to refract and reflect, forming a localised standing wave pattern concentrated between the two footpoints of a closed loop. The refraction process induces a ponderomotive force, a non-linear (i.e. not directly proportional to wave amplitude) force that is directed upwards and acts selectively on ions. Since more low-FIP elements are ionised in the chromosphere and transition region, they are more strongly affected than high-FIP elements, which remain mostly neutral. This selective upward acceleration increases the abundance of low-FIP ions in the corona relative to their photospheric values, producing the observed coronal FIP bias~\cite{2015LRSP...12....2L}.

While elements are typically grouped into high-FIP and low-FIP elements depending on whether their FIP is above or below 10~eV, not all elements fit cleanly into this classification. Sulphur (FIP = 10.36~eV) lies just above this boundary, and has conventionally been classified and observed as a high-FIP element. For example, remote sensing observations from EIS routinely suggest S behaves similarly to other high-FIP elements in its abundance~\cite{Brooks2011ApJ...727L..13B,2013ApJ...778...69B,Baker2015,Baker2018,To2023ApJ...948..121T}. However, recent EIS studies have shown that S can exhibit both high- and low-FIP behaviour depending on the local magnetic and plasma conditions. For example, variable behaviour of S has been observed in flares or AR coronal rain~\cite{To2021ApJ...911...86T,Brooks2024ApJ...962..105B,To2024A&A...691A..95T, Ng2026arXiv260323098N}, complex ARs with strong light bridges with coalescing or merging sunspot umbrae~\cite{Baker2024ApJ...970...39B}, pre-existing loops neighbouring emerging loops~\cite{Mihailescu2023ApJ...959...72M}, \rev{and even in coronal plumes observed by Solar Orbiter~\cite{Mzerguat2026arXiv260223170M}}. Likewise, in-situ solar wind measurements often observe S enriched relative to photospheric values, consistent with the low-FIP behaviour~\cite{2016SSRv..201...55A,Brooks2017Aug,2020ApJ...895...36K}. \rev{Recent work has also suggested that the Si/S abundance ratio varies with open versus closed magnetic field and can be used as a diagnostic for solar energetic particles source regions~\cite{2018SoPh..293...47R, 2021SciA....7...68B, 2024ApJ...976..152Y}.} These results suggest that S represents a unique boundary case whose behaviour that varies across solar structures, providing an opportunity to use S as a sensitive probe of the effects of different phenomena on the fractionation process.

Here, we interpret Sulphur's variable behaviour using the ponderomotive force model. This model suggests that fractionation is heavily dependent on magnetic field strength and whether the Alfv\'{e}n waves are in resonance or off-resonance with coronal loops. Stronger magnetic fields change the height of the $\beta = 1$ layer (where gas and magnetic pressures are equal) in the chromosphere, while loop length may affect Alfv\'{e}n wave resonance. Consequently, S may behave as a high- or low-FIP element under different magnetic conditions, making it an ideal element for studying magnetic influences on abundance patterns. To systematically test this, we examine how relative FIP biases, with and without S, correlate with both magnetic field strength and loop length across a set of EIS rasters. Previous studies have examined the connection between magnetic field strength and FIP bias using absolute magnetic flux densities obtained directly from Helioseismic and Magnetic Imager (HMI;~\cite{2012SoPh..275..207S}) magnetograms with somewhat inconsistent results. Mihailescu et al.~\cite{Mihailescu2022ApJ...933..245M} examined 28 ARs of various types and evolutionary stages \rev{using the Si/S diagnostic}, finding that FIP bias generally increased with average magnetic flux density up to $\sim$200~G, and that as ARs evolve from higher to lower flux density during decay, the FIP bias correspondingly decreases with time. In contrast, Baker et al.~\cite{2013ApJ...778...69B} found a \rev{weak positive }correlation ($r = 0.37$), with the FIP bias trend continuing up to 500~G without evidence of plateauing. These differing results highlight the need for a more sophisticated approach when interpreting abundance trends in relation to magnetic properties. Here we advance beyond the previous work to quantify the relationship between observed FIP bias and structural properties, including loop length and mean magnetic field along a coronal loop by using potential field source surface extrapolations (PFSS) to provide a more comprehensive assessment of how loop properties influence elemental fractionation. \rev{This study focuses on a single AR (NOAA AR~11967) observed over five days using nine EIS rasters.}



\section{Observations and Data Analysis}

\begin{figure}[!t]
    \centering
    \includegraphics[width=\linewidth]{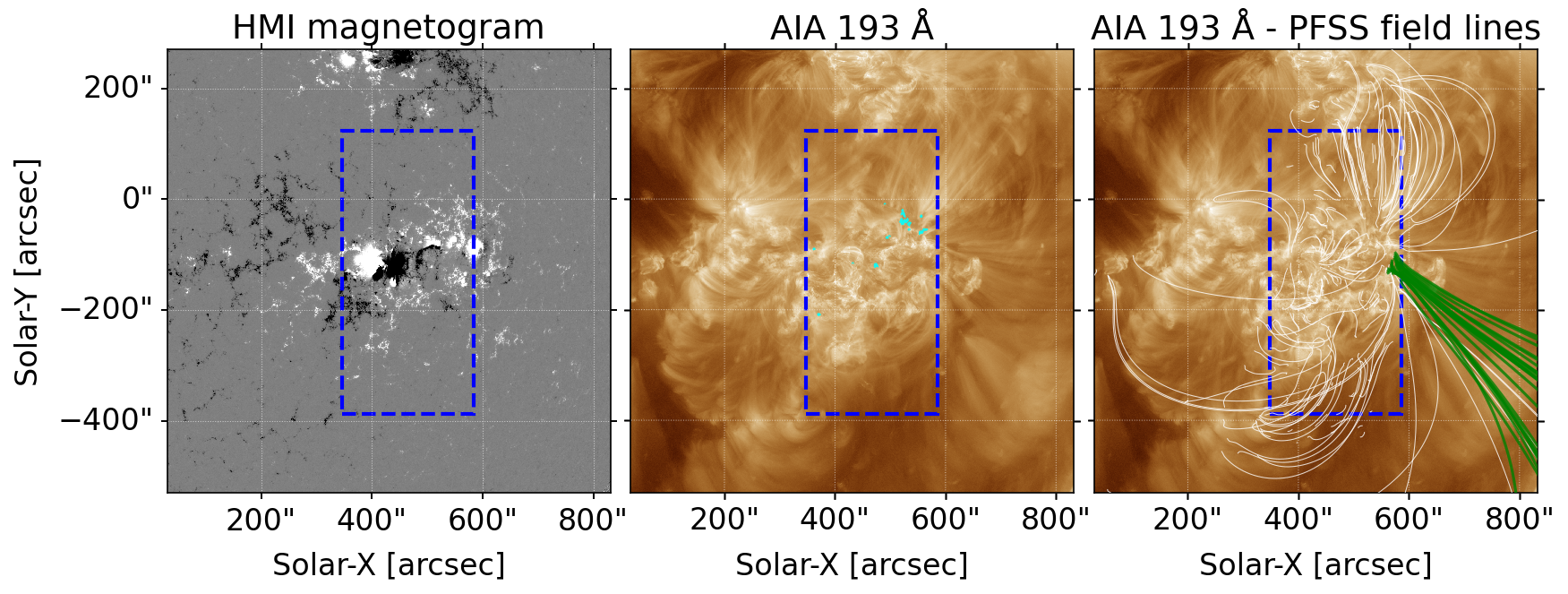}
    \caption{\rev{Overview of AR~11967 on 2014-02-05T10:41. Left to right: HMI line-of-sight magnetogram, AIA 193~\AA\ image, and AIA 193~\AA\ image with PFSS-extrapolated field lines overlaid. The EIS field of view (FOV) is outlined in blue. Cyan contours in the middle panel mark EIS Fe XII 195.12 Doppler velocity $(\le -10\, \mathrm{km\,s^{-1}})$ within the EIS FOV. Closed PFSS field lines are shown in white and open field lines in green.}}
    \label{fig:HMI_AIA_PFSS}
\end{figure}

Our focus is on AR~11967, a highly complex region \rev{(Hale class $\beta\gamma\delta$ during 2014 Feb 1--5~\cite{NOAA_SRS_20140202,NOAA_SRS_20140203,NOAA_SRS_20140204,NOAA_SRS_20140205,NOAA_SRS_20140206})} that was visible on the southern hemisphere of the Sun from 2014 January 31 to 2014 February 7. An overview of AR~11967 and its magnetic topology is shown in Figure~\ref{fig:HMI_AIA_PFSS}. \rev{Additional AIA 193~\AA\ and PFSS overlay panels for all nine rasters analysed in the study are shown in the Appendix Figure~\ref{fig:All_AIA_PFSS}.} \rev{The PFSS extrapolation shows that the open field lines lie predominately outside the EIS FOV, while the Fe XII~195.12~\AA\ Doppler velocity upflow contours within the FOV show little spatial overlap with the open field lines (Figure~\ref{fig:HMI_AIA_PFSS} middle and last panel).} \rev{This region exhibited frequent flux emergence and reconnection, creating a range of interesting solar events (To et al.~\cite{To2021ApJ...911...86T}), including evolving sunspot umbrae separated by light bridges.} The continuous Hinode/EIS coverage of AR~11967 during its disk passage provides a rich dataset of nine rasters obtained between 2014 February 1 and 2014 February 5, with multiple strong spectral lines suitable for FIP bias analysis. To quantify the S fractionation, we can compare the behaviour of 4 diagnostic line pairs, including Si\,\textsc{x} 258.37~\AA\ / S\,\textsc{x} 264.23~\AA, S\,\textsc{xi} 188.68~\AA\ / Ar\,\textsc{xi} 188.81~\AA, Ca\,\textsc{xiv} 193.87~\AA\ / Ar\,\textsc{xiv} 194.40~\AA, and Fe\,\textsc{xvi} 262.98~\AA\ / S\,\textsc{xiii} 256.69~\AA. These line pairs have been used in previous studies~\cite{Baker2019ApJ...875...35B,Baker2024ApJ...970...39B, To2024A&A...691A..95T}. Each line pair has similar formation temperatures and minimal density sensitivity, ensuring robust abundance comparison. Figure~\ref{fig:emissivities_ratio} in the Appendix shows the emissivity and ratio curves for each line pair, highlighting their suitability for FIP bias analysis. 

\subsection{Differential Emission Measure and Composition Maps}

\begin{table}[!t]
    \caption{Hinode/EIS study details used in this study}
    \centering
    \begin{tabular}{ll}
    \hline
    Study Number & 437 (HPW021\_VEL\_240x512v1), 403 (Atlas\_30) \\
    Emission Lines & \begin{tabular}[t]{l}
    \textbf{Composition diagnostics:} \\
    Ca XIV 193.87~\AA, Ar XIV 194.40~\AA \\
    Fe XVI 262.98~\AA, S XIII 256.69~\AA \\
    Si X 258.38~\AA, S X 264.23~\AA \\
    S XI 188.68~\AA, Ar XI 188.81~\AA \\
    \textbf{DEM lines:} \\
    Fe VIII 185.21~\AA, Fe VIII 186.60~\AA \\
    Fe IX 188.50~\AA, Fe IX 197.86~\AA \\
    Fe X 184.54~\AA, Fe XI 188.22~\AA \\
    Fe XI 188.30~\AA, Fe XII 186.88~\AA \\
    Fe XII 192.39~\AA, Fe XII 195.12~\AA \\
    Fe XIII 202.04~\AA, Fe XIII 203.83~\AA \\
    Fe XIV 264.79~\AA, Fe XIV 270.52~\AA \\
    Fe XV 284.16~\AA, Fe XVI 262.98~\AA \\
    Fe XVII 254.87~\AA, Fe XXIII 263.76~\AA \\ 
    Fe XXIV 255.10~\AA \\
    \end{tabular} \\
    Field of View & 240$^{\prime\prime}$ $\times$ 512$^{\prime\prime}$ (437) \\
                  & 120$^{\prime\prime}$ $\times$ 160$^{\prime\prime}$ (403) \\
    Rastering & 1$^{\prime\prime}$ slit, 120 positions, 2$^{\prime\prime}$ coarse steps (437) \\
              & 2$^{\prime\prime}$ slit, 60 positions, 2$^{\prime\prime}$ steps (403) \\
    Exposure Time & 60 s (437), 30 s (403) \\
    Total Raster Time & 2 hr (437), 30 minutes (403) \\
    Reference Spectral Window & Fe XII 195.12~\AA \\
    \hline
    \end{tabular}
    \label{tab:EIS_details}
\end{table}

\begin{figure}[htbp]
    \centering
    \includegraphics[width=\linewidth]{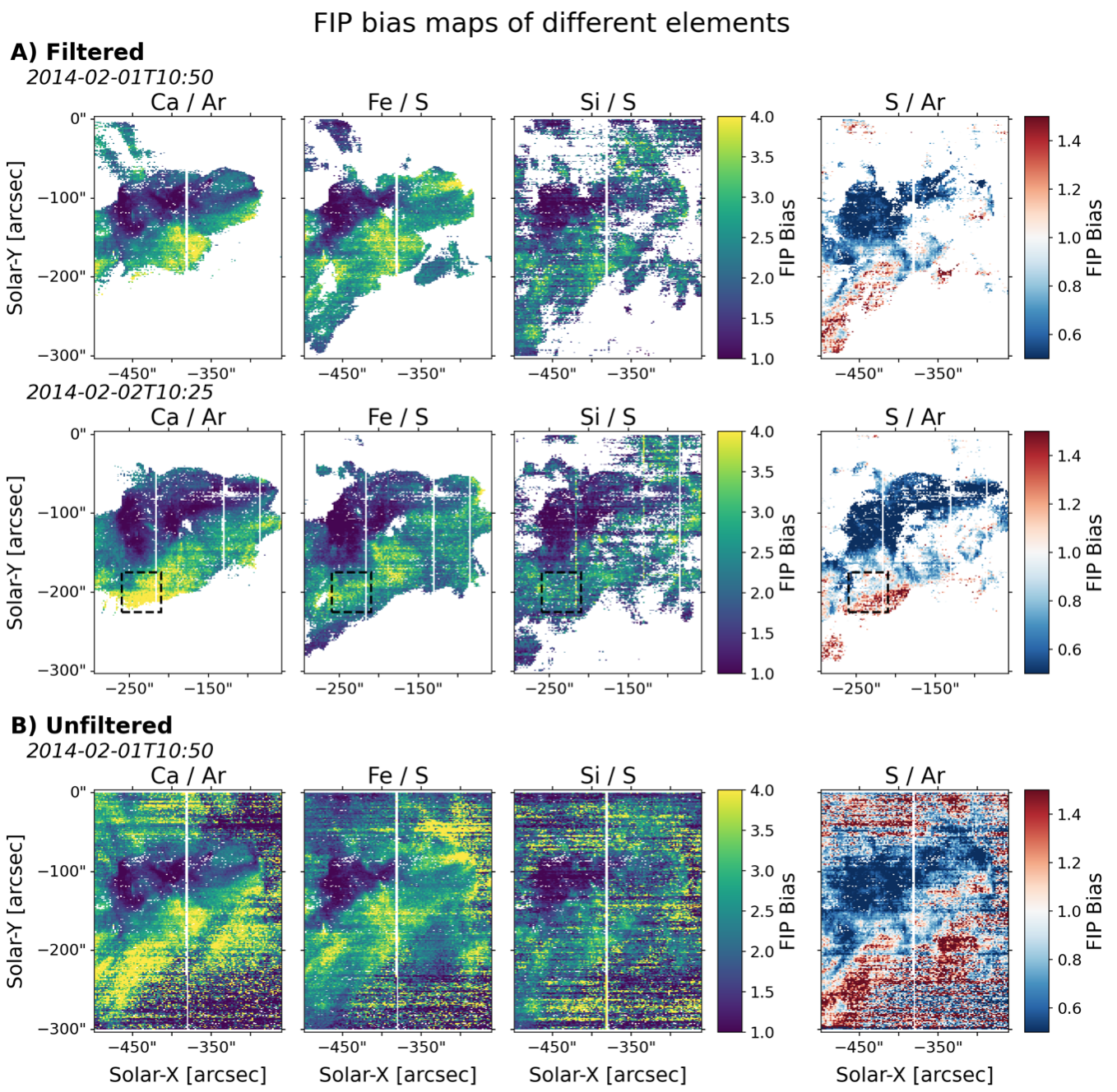}
    \caption{\rev{Sample} FIP bias maps for the four diagnostic line pairs, derived from DEM-corrected intensities during nine EIS observations in February 2014. Section A shows the filtered maps with both $\chi^2$ and intensity masking applied for each observation date shown: 2014-02-01T10:50, 2014-02-02T10:25. Section B shows the unfiltered map for 2014-02-01T10:50. \rev{The full analysis consists of 9 rasters spanning from 2014-02-01 to 2014-02-05 (see Appendix Figure~\ref{fig:All_FIP_PFSS}).} For the Ca/Ar, Fe/S, and Si/S, blue regions indicate near-unity ratios (unenhanced abundance), while yellow regions show enhancement of the low-FIP element; for S/Ar, blue (red) shows regions where S is depleted (enhanced) relative to Ar. These maps are used to evaluate the FIP behaviour of S in comparison to high- and low-FIP reference elements.}
    \label{fig:Composition_Maps}
\end{figure}

To obtain the FIP bias maps, we used a similar approach to previous EIS FIP bias studies~\cite{Brooks2011ApJ...727L..13B,Baker2013Nov}. We first fitted each spectral line profile. Emissions from the consecutive ionization stages of Fe VIII -- Fe XXIV were fitted with a single Gaussian function, with the exception of Fe XI 188.22~\AA, Fe XI 188.30~\AA, Fe XIII 203.83~\AA, Fe XIV 270.52~\AA, Fe XV 284.16~\AA, and Fe XXIV 255.10~\AA, which were fitted with multiple Gaussian profiles. Following this, we calibrated the intensity of each spectral line according to the Warren et al. 2014 calibration~\cite{Warren2014ApJS..213...11W}\footnote{The calibration used here is available at \url{https://github.com/andyto1234/EISPAC-Tutorial___Calibrations}, which reproduces the \texttt{SolarSoftWare (SSW)} \texttt{IDL} calibration in \texttt{Python}.}. This corrects for the long-term, wavelength-dependent sensitivity degradation of EIS. Then, we calculated the DEM using exclusively Fe lines (Fe VIII -- Fe XXIV) as shown in table~\ref{tab:EIS_details}. The DEM was then calculated from the fitted intensities using the regularized inversion technique~\cite{2012A&A...539A.146H}, with emissivity from CHIANTI v10.0.1~\cite{Dere1997Oct, DelZanna2021ApJ...909...38D}, photospheric abundances from Scott et al.~\cite{Scott2015A&A...573A..25S, Scott2015bA&A...573A..26S}, and electron density derived separately using the Fe XIII 202.04~\AA/Fe XIII 203.83~\AA\ ratio. Pixels with a $\chi^2$ larger than the number of Fe lines used for the DEM calculation were filtered out. To calculate the FIP bias, we first scaled the DEM to match the intensity of the `numerator' element's emission line (Ca, Fe, Si or S), with the FIP bias then determined from the ratio of predicted to observed intensities of the `denominator' line (Ar or S). This method is designed to remove the temperature and density effects for a robust calculation of the FIP bias. Fitting of all of the EIS spectral lines was done using the~\texttt{eispac} python package~\cite{Weberg2023JOSS....8.4914W}. 
Figure~\ref{fig:Composition_Maps} shows FIP bias maps formed using 4 composition diagnostic ratios across different dates, each formed at different temperatures. From left to right, we have Ca XIV~193.87~\AA/Ar XIV~194.40~\AA, Fe XVI~262.98~\AA/S XIII~256.69~\AA,  Si~X~258.38~\AA/S X~264.23~\AA, and S~XI~188.68~\AA/Ar XI~188.81~\AA. \rev{We use a different colour table for S/Ar as this diagnostic compares a mid-FIP element with a high-FIP element.} To further eliminate the effect of noise from the FIP bias, we applied an intensity-based filter to the composition diagnostics. For each EIS raster, we masked pixels whose intensities fell below empirically chosen, emission-line-specific thresholds designed to exclude low-signal regions. The thresholds were applied uniformly across both small- and large-FOV rasters (for the numerator/denominator lines respectively, we required Ca XIV $\geq 1.2 $ and Ar XIV $ \geq 0.5~\mathrm{erg}\,\mathrm{cm}^{-2}\,\mathrm{s}^{-1}\,\mathrm{sr}^{-1} $; Fe XVI $ \geq 3 $ and S XIII $ \geq 5~\mathrm{erg}\,\mathrm{cm}^{-2}\,\mathrm{s}^{-1}\,\mathrm{sr}^{-1} $; Si X $ \geq 2 $ and S X $ \geq 0.8~\mathrm{erg}\,\mathrm{cm}^{-2}\,\mathrm{s}^{-1}\,\mathrm{sr}^{-1} $; S XI $ \geq 0.9 $ and Ar XI $ \geq 0.4~\mathrm{erg}\,\mathrm{cm}^{-2}\,\mathrm{s}^{-1}\,\mathrm{sr}^{-1} $), ensuring that FIP bias was only computed in regions with strong emission at temperatures relevant to the diagnostic lines used (e.g., log$T/K$ = 6.6 for Ca/Ar). Figure~\ref{fig:Composition_Maps} section A shows \rev{some of} the intensity filtered \rev{composition} maps, \rev{representing 2 out of 9 rasters analysed in this study}, while section B shows the unfiltered composition map as a reference for our method. For the Ca/Ar, Fe/S, and Si/S diagnostics, blue regions indicate near-unity relative FIP bias, where neither element is preferentially enhanced, while brighter colours such as yellow mark locations where low-FIP elements are more abundant relative to their high-FIP counterparts. In contrast, for S/Ar, blue (red) shows regions where S is depleted (enhanced) relative to Ar. 

Since the Ca/Ar composition diagnostic is the only ratio that does not involve the uncertainty associated with the variable abundance behaviour of S, we use this line pair as a rough baseline reference for abundance enhancement. When comparing the FIP bias behaviour of Fe/S and Ca/Ar FIP bias (log~T/K$\sim$6.4 and 6.6, respectively), they share broadly similar patterns across multiple rasters, particularly in areas such as the core of the AR, where the FIP bias values are both enhanced. Similarly, Si/S, which is sensitive to cooler plasma temperatures (log~T/K$\sim$6.2), shares a broadly similar behaviour to Ca/Ar across all observations. However, in certain locations, it can be seen that both the Fe/S and Si/S FIP biases deviate from Ca/Ar. For example, in the 2014-02-02T10:25 row at [x, y] $\sim$ [-235\arcsec, -200\arcsec] (black dashed box in Figure~\ref{fig:Composition_Maps}), the Ca/Ar FIP bias is strongly enhanced, whereas Fe/S and Si/S remain close to intermediate values of $\sim$ 2--2.5, with some patches approaching $\sim$3. By contrast, S/Ar consistently reaches values of $\sim$ 1.3--1.5, indicating that S is enhanced relative to Ar and may not be behaving uniformly as a high-FIP element in this region, diverging from both Ca/Ar and the S-involving diagnostics.

\subsection{Potential Field Source Surface Extrapolation}

To further understand the fractionation of S, and the magnetic environments prone to its fractionation, we compared the behaviour of S against coronal loop length and mean magnetic flux density of loops using a PFSS extrapolation. The PFSS method enables tracing of field lines from the solar photosphere out to a source surface at $\sim$2.5~R$_\odot$, providing a 3D representation of magnetic connectivity and structure that can be compared with spatial variations in elemental composition. The PFSS solutions were computed using the~\texttt{PFSSpy} package~\cite{2020JOSS....5.2732S}, adopting synoptic HMI magnetograms as the lower boundary. These maps are assembled from the central-meridian full-disk observations taken over each Carrington rotation ($\sim$27 days) and are updated as new HMI magnetograms become available, with the central meridian region replaced by the most recent observations~\cite{2015SoPh..290.1507H}. This approach provides a high spatial resolution, and timely boundary condition near the Earth-facing side of the Sun. Since our analysis focuses on statistical correlations within the EIS FOV, the updated synoptic HMI maps provide a consistent lower boundary for PFSS without requiring additional flux transport or assimilation modelling. PFSS is a commonly used method for approximating the large-scale solar coronal magnetic field under the assumption that the corona is current-free ($\nabla \times \mathbf{B} = 0$) and that the magnetic field becomes radial at a designated outer boundary (the source surface). Figure~\ref{fig:HMI_AIA_PFSS} shows one of the examples of PFSS field lines overlaid on AIA 193~\AA\ image on 1 February 2014\rev{, while Figure~\ref{fig:AIA_PFSS_above150g_Appendix} in the Appendix presents the corresponding PFSS field lines for all nine EIS rasters. Overall, the PFSS field lines show good correspondence with the coronal structures seen in the AIA 193~\AA\ images.} \rev{To enable pixel by pixel comparison, the EIS Fe~XII 195.12~\AA\ intensity map were co-aligned to the closest time AIA 193~\AA\ image using cross-correlation technique in \texttt{Python}, and the HMI based PFSS seed locations were mapped onto the EIS pixel grid using the map's world coordinate system (WCS).} Although PFSS extrapolation makes a current-free assumption, our focus is on the statistical correlations between magnetic loop properties and elemental composition across several days and multiple rasters, for which this approximations is sufficient for our purposes.

\begin{figure}
    \centering
    \includegraphics[width=\linewidth]{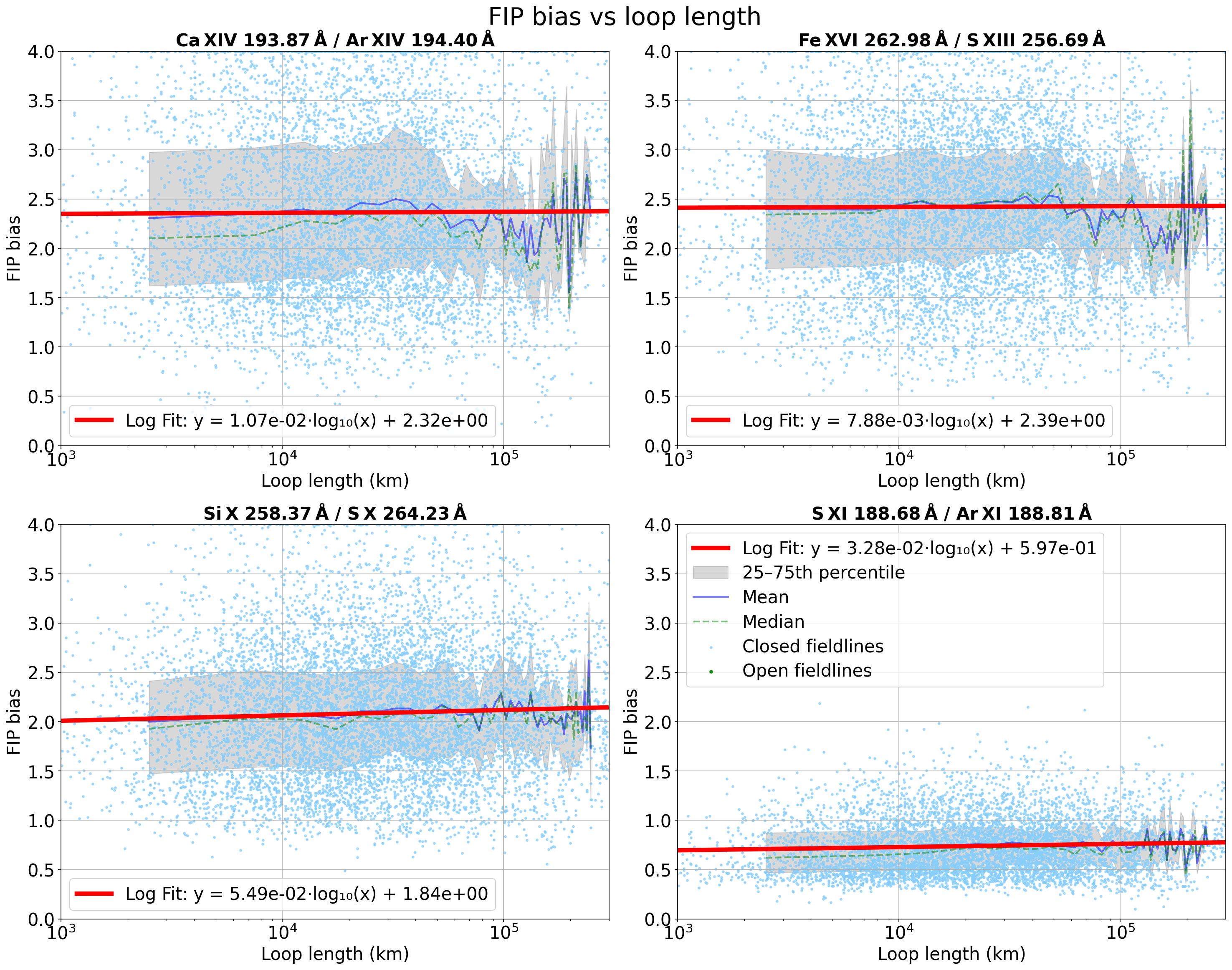}
    \caption{Elemental abundance versus magnetic loop length for all nine rasters combined, shown separately for diagnostic line pair. Blue points represent closed field lines; green dashed line indicates the median abundance per bin, calculated using a 5,000~km binning of loop length; blue solid line shows the corresponding mean; gray shaded region indicates the 25th -- 75th percentile range. A logarithmic fit is plotted in red, with its equation displayed on the left of each graph. The exact number of valid pixels are: Ca/Ar ($N = 7424$), Fe/S ($N = 8266$), Si/S ($N = 8789$), S/Ar ($N = 6373$).}
    \label{fig:Abundance_length}
\end{figure}

For each EIS raster, PFSS field extrapolations were initiated from seed points (starting points) defined from HMI pixels within the EIS FOV that exceed the magnetic field thresholds. Pixels with field strength between $-5$ to $+5$~G were excluded to remove regions of very weak magnetic field. Field lines were traced from each seed point through the resulting PFSS solution with a step size of $0.01$~$R_\odot$ and a maximum of 70,000 steps, until they reached either the source surface (2.5~$R_\odot$) or returned to the solar surface. The resulting field line coordinates were then used for loop length and mean magnetic field calculation.

\subsection{Abundance versus Magnetic Field Properties}

To investigate how loop properties affect elemental composition, we associated each EIS pixel with two quantities derived from the PFSS-traced field lines: 1) total loop length calculated by summing the three-dimensional segment distances along each traced field line, and 2) mean magnetic field strength computed by averaging the magnitude of the magnetic field vector at each traced point. Both quantities were then benchmarked against the FIP biases in the composition maps. The final comparison uses data combined from all nine rasters to improve statistical significance.

Figure~\ref{fig:Abundance_length} shows the relationship between loop length and FIP bias for all four FIP bias diagnostics. Blue points indicate individual closed field lines, while the blue solid and green dashed lines show the mean and median abundances within each 5,000~km loop length bin, respectively. \rev{We note that only a small number of open field lines were found within the EIS FOV; in PFSS these are defined as field lines that reach the 2.5~R$_\odot$ source surface and therefore populate the extreme long loop length tail (see Figure~\ref{fig:All_AIA_PFSS} in the Appendix for the location of open field lines in all rasters). For clarity, we restrict the x-axis range in Figure~\ref{fig:Abundance_length} to the loop length range containing the majority of well populated bins, since including the few open field points would disproportionately expand the x-axis and obscure the dominant closed field distribution.} Across all four composition diagnostics, the fitted slopes range 0.008 -- 0.055, indicating either a very weak or flat trend across all panels. This shows that FIP bias shows no correlation with loop length.

\begin{figure}
    \centering
    \includegraphics[width=\linewidth]{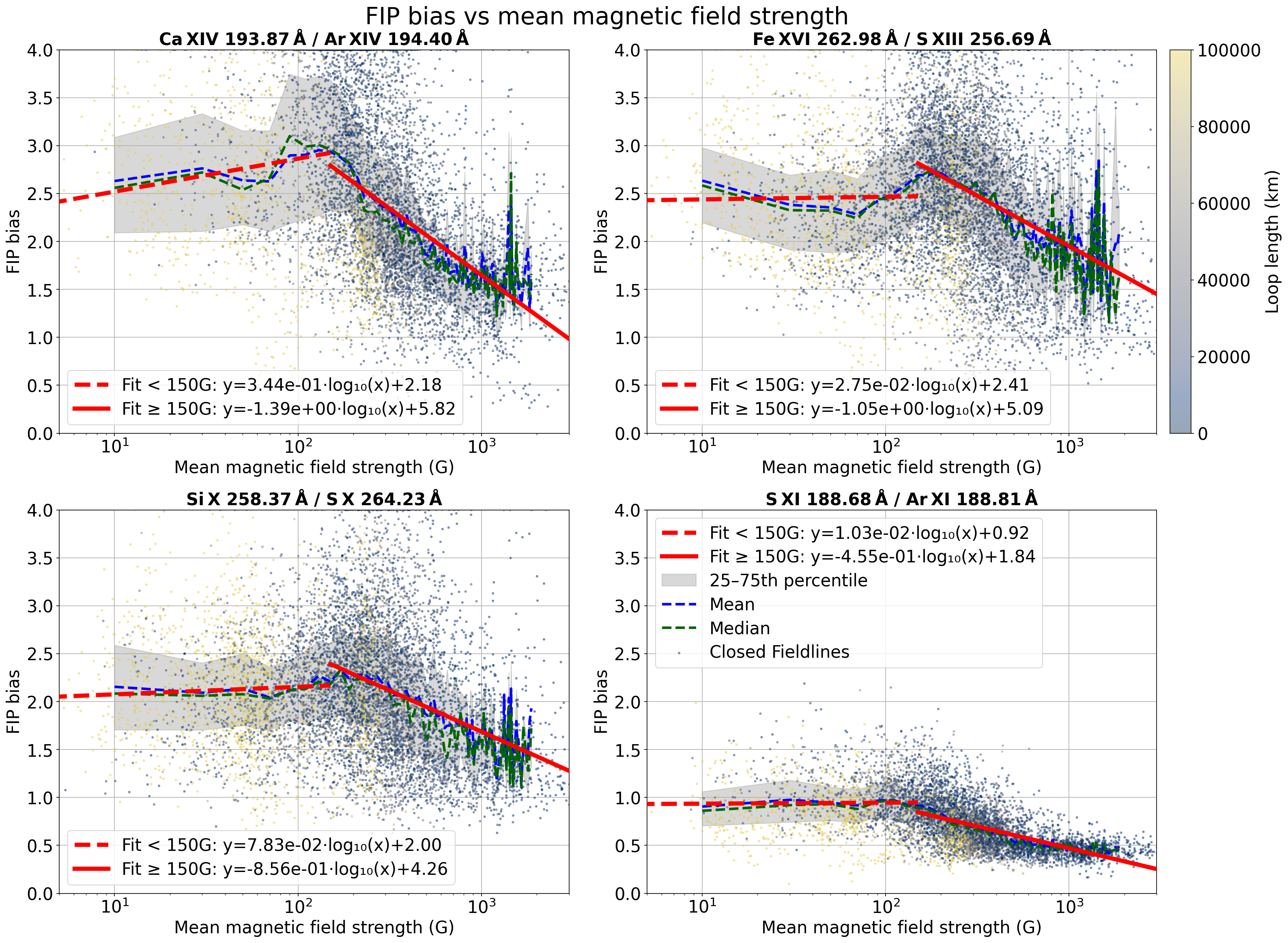}
    \caption{Elemental abundance versus mean magnetic field strength along PFSS-traced field lines for all nine rasters combined, shown separately for each diagnostic line pair. The colour of dots indicates loop length, from shorter loops in blue to longer loops in yellow; green dashed line indicates the median abundance per bin, calculated using a 20~G binning of magnetic field strength; the blue dashed line shows the corresponding mean; gray shaded region indicates the 25th -- 75th percentile range. Red dashed and solid lines show fitted trends below and above 150~G, with their fitted equations displayed on the left of each graph. The exact number of valid pixels (< 150~G /$\geq$ 150~G) are: Ca/Ar (1642, 5757), Fe/S (2325, 5916), Si/S (3037, 5737), S/Ar (1773, 4598); See Figure~\ref{fig:histogram} for the corresponding pixel count distribution.}
    \label{fig:Abundance_mean_magnetic}
\end{figure}

Figure~\ref{fig:Abundance_mean_magnetic} shows the relationship between mean magnetic field strength within a loop and its FIP biases. For comparison, Mihailescu et al.~\cite{Mihailescu2022ApJ...933..245M} reported that the positive trend observed below $\sim$200~G does not appear to continue at higher field strengths, but they did not establish a systematic decrease. An interesting finding is that the FIP bias/magnetic field strength slope between each FIP bias diagnostic varies depending on the element involved in the FIP bias calculation, showing differences between diagnostics with and without S. Ca/Ar FIP bias shows a sharper slope beyond a mean magnetic field strength of 150~G, while Si/S and Fe/S both show a shallower slope after this threshold. Combining this with the S/Ar composition diagnostics indicates that S is being depleted in loops with high magnetic field strength.

To quantify these relationships, we fitted two log-linear lines below and above 150~G to relate the relative FIP bias of different diagnostics to the mean magnetic field strength (Red dashed and solid lines in Figure~\ref{fig:Abundance_mean_magnetic}), with the results given as:

\begin{align}
A_{\mathrm{Ca/Ar}} &=
\begin{cases}
    0.34 \cdot \log_{10}(B) + 2.18, & \text{if } B < 150~\mathrm{G} \\
    -1.39 \cdot \log_{10}(B) + 5.82, & \text{if } B \geq 150~\mathrm{G}
\end{cases} \\
A_{\mathrm{Fe/S}} &=
\begin{cases}
    0.028 \cdot \log_{10}(B) + 2.41, & \text{if } B < 150~\mathrm{G} \\
    -1.05 \cdot \log_{10}(B) + 5.09, & \text{if } B \geq 150~\mathrm{G}
\end{cases} \\
A_{\mathrm{Si/S}} &=
\begin{cases}
    0.078 \cdot \log_{10}(B) + 2.00, & \text{if } B < 150~\mathrm{G} \\
    -0.86 \cdot \log_{10}(B) + 4.26, & \text{if } B \geq 150~\mathrm{G}
\end{cases} \\
A_{\mathrm{S/Ar}} &=
\begin{cases}
    0.01 \cdot \log_{10}(B) + 0.92, & \text{if } B < 150~\mathrm{G} \\
    -0.46 \cdot \log_{10}(B) + 1.84, & \text{if } B \geq 150~\mathrm{G}
\end{cases}
\end{align}

\rev{Below 150~G, the fitted slopes relating FIP bias to mean magnetic field are very small (0.01 -- 0.34), indicating little change in FIP bias with increasing mean magnetic field.} However, above 150~G, both Fe/S and Si/S diagnostics show a similar slope $\sim$-0.86 -- -1.05, whereas Ca/Ar, with no influence of S, has a much sharper slope at -1.39. This result is further verified through the S/Ar diagnostic, showing a near flat slope $\sim$0.01 that decreases to $\sim$-0.46 beyond 150~G, signifying that S is being depleted in loops with a high mean magnetic field strength.
\begin{figure}
    \centering
    \includegraphics[width=0.7\linewidth]{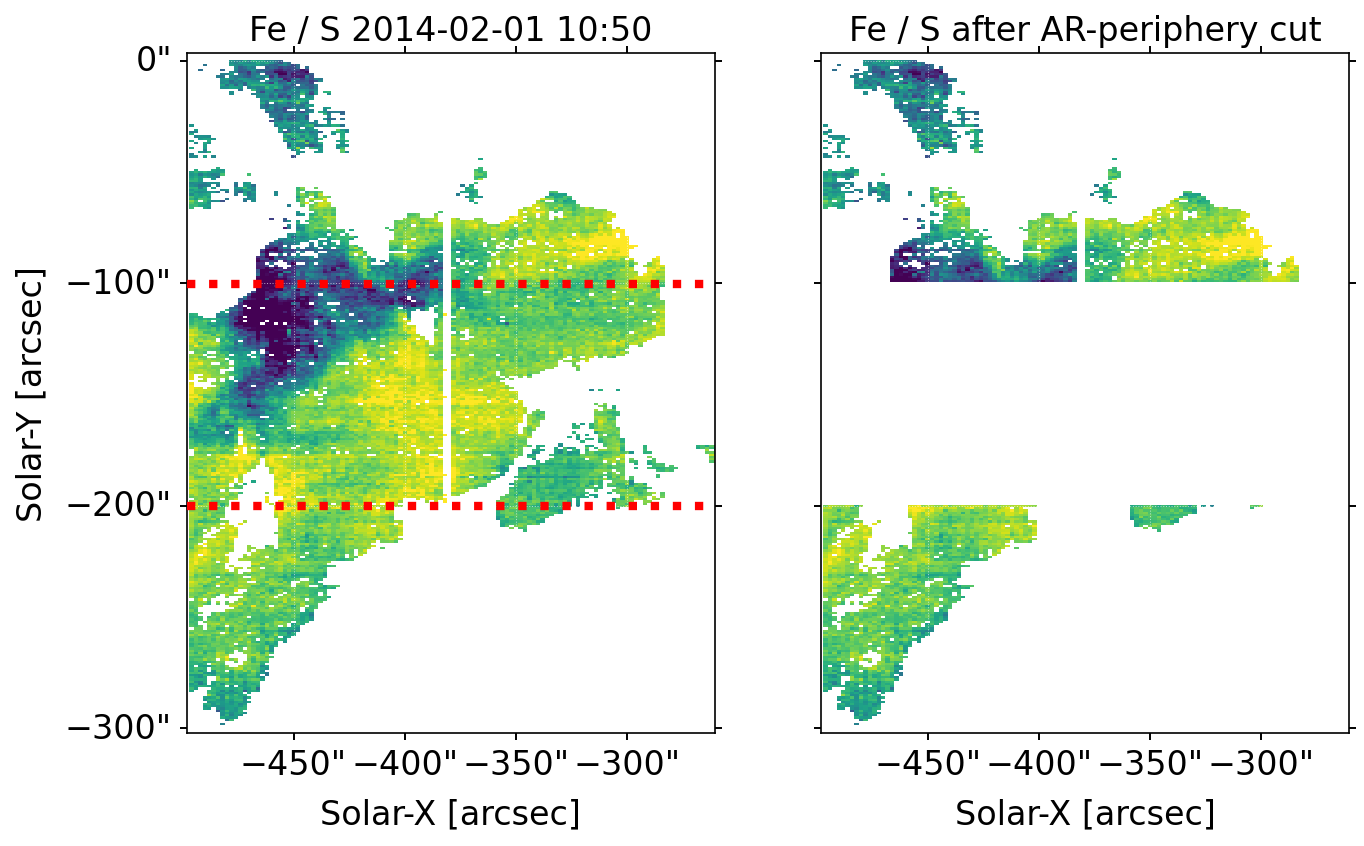}
    \caption{\rev{Example of the Solar-Y mask used for the AR periphery analysis in~\ref{fig:refcomment_Abundance_length} and~\ref{fig:refcoment_Abundance_mean_magnetic}. Left: cleaned Fe/S composition map before applying the mask. Right: the same map after excluding the central band $-200" < Y < -100"$.}}
    \label{fig:AR_periphery_mask}
\end{figure}

\begin{figure}
    \centering
    \includegraphics[width=\linewidth]{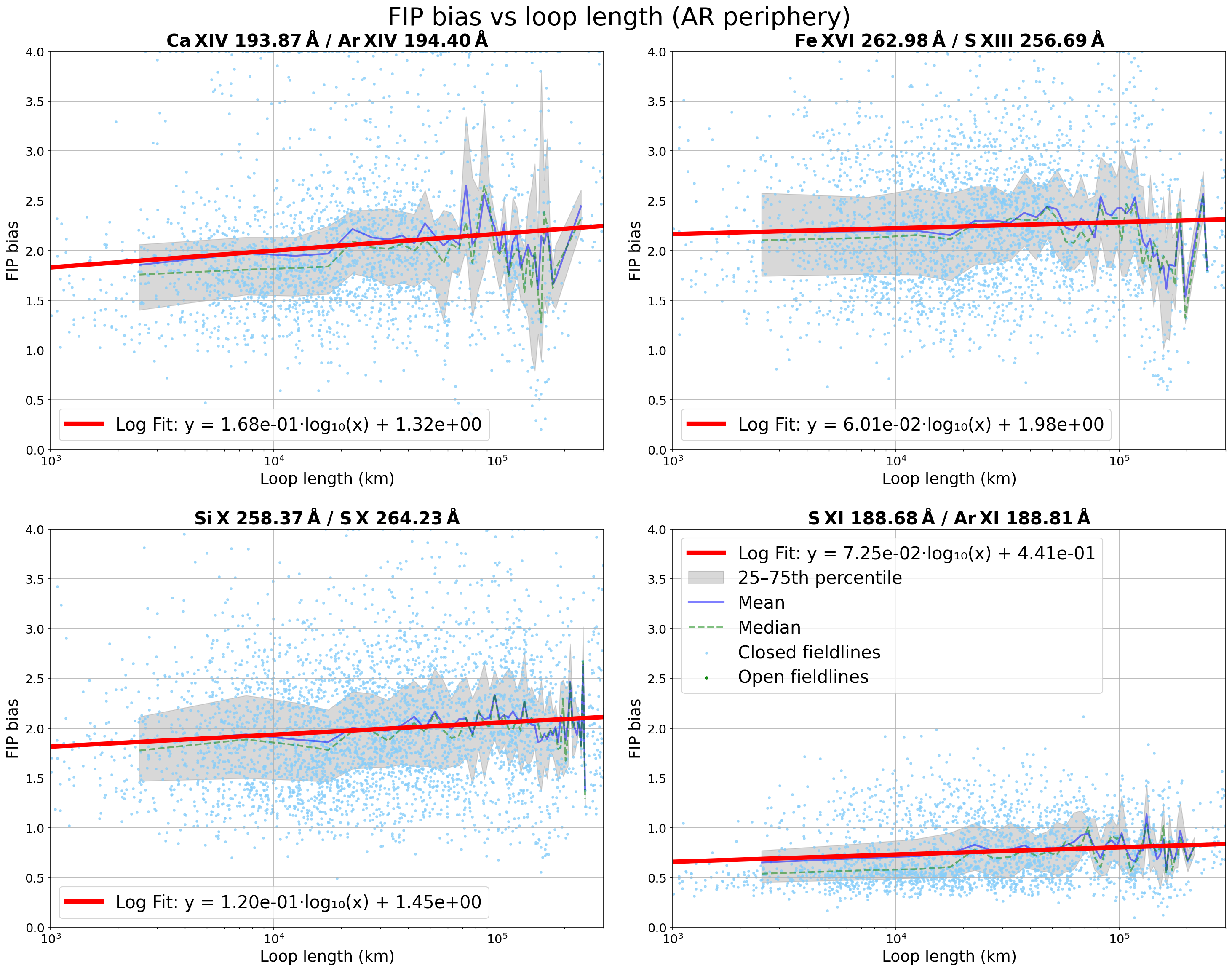}
    \caption{\rev{Same as Figure~\ref{fig:Abundance_length}, except repeated for the AR periphery of the EIS FOV, excluding the bright, magnetically complex core. The exact number of valid pixels are: Ca/Ar ($N = 2554$), Fe/S ($N = 3094$), Si/S ($N = 3947$), S/Ar ($N = 2494$).}}
    \label{fig:refcomment_Abundance_length}
\end{figure}

\begin{figure}
    \centering
    \includegraphics[width=\linewidth]{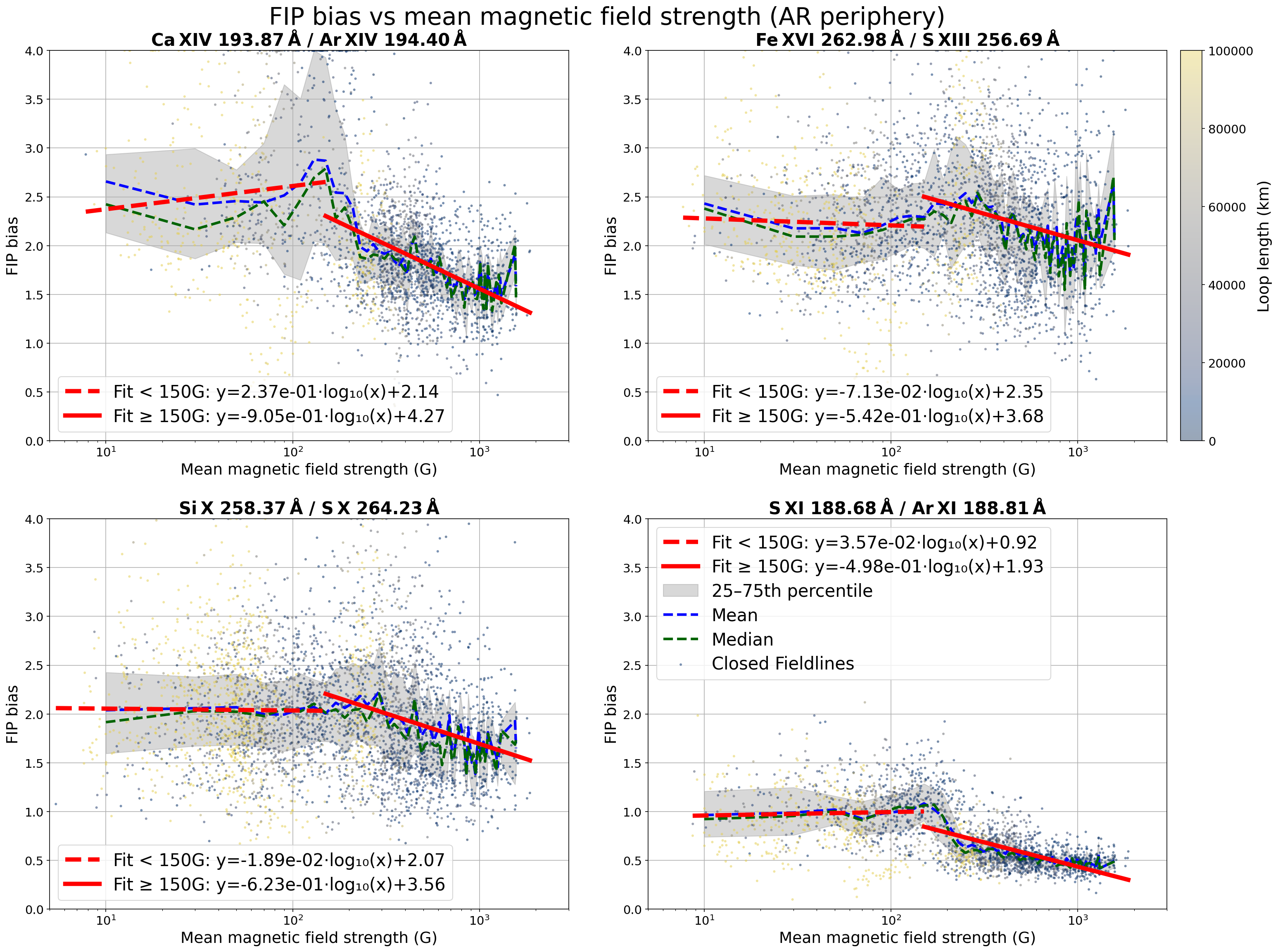}
    \caption{\rev{Same as Figure~\ref{fig:Abundance_mean_magnetic}, except repeated for the AR periphery of the EIS FOV, excluding the bright, magnetically complex core. The exact number of valid pixels (< 150~G /$\geq$ 150~G) are: Ca/Ar (639, 1913), Fe/S (1184, 1908), Si/S (1921, 2025), S/Ar (946, 1547).}}
    \label{fig:refcoment_Abundance_mean_magnetic}
\end{figure}

\rev{Understanding PFSS extrapolations may not fully capture the magnetically complex core of the active region, we repeated both the loop length and mean magnetic field strength analyses using only the AR periphery of each raster. The periphery was defined using the simple Solar-Y mask shown in Figure~\ref{fig:AR_periphery_mask}, which excludes the central band of the EIS FOV and therefore reduces the contribution from the complex AR core. This masking reduces the number of valid pixels by about 2.5--2.6 times. Nevertheless, the overall results remain unchanged. The new loop length analysis in Figure~\ref{fig:refcomment_Abundance_length} shows the same weak or absent dependence of FIP bias on loop length, with only modest changes in slope, while the mean magnetic field strength analysis (Figure~\ref{fig:refcoment_Abundance_mean_magnetic}) preserves the same piecewise behaviour: weak trends below 150~G and a clear decrease above 150~G, although with slightly shallower slopes in the high-field regime. One possible explanation is that PFSS extrapolations neglect electric currents. In a non-potential field, currents introduce shear and twist that lengthen field lines, which would reduce the mean magnetic field strength sampled along a loop and could therefore flatten the high-field trend.}



\section{Discussion and Conclusion}

In this study, we compared four FIP bias diagnostics (Ca/Ar, Fe/S, Si/S, and S/Ar) with 1) loop length and 2) average magnetic field strength derived from PFSS extrapolations, using nine Hinode/EIS rasters spanning five days of observations of AR~11967. We constructed FIP bias maps using differential emission measure (DEM) analysis and correlated these with loop length and loop mean magnetic field strength from PFSS extrapolations. 

We find that across all four composition diagnostics, FIP bias shows weak to no correlation with loop length. Instead, we identify a significant relationship with the mean magnetic field strength of loops. Specifically, FIP bias remains relatively constant below 150~G but systematically decreases with increasing field strength above this threshold across all four diagnostics. This extends previous observational results~\cite{Mihailescu2022ApJ...933..245M}, which showed an increase in FIP bias up to $\sim$200~G followed by a flattening, by demonstrating that the high-field regime is characterised by a clear decline. However, the magnitude of this trend differs significantly among diagnostics. Ca/Ar shows the steepest negative slope above 150~G, while Fe/S and Si/S exhibit moderately steep decreases, with S/Ar showing the shallowest reduction.

\rev{An interesting question is whether the apparent elbow near 150~G has a physical origin, or instead reflects limitations of the magnetic extrapolation. One possible explanation is that loops with mean field strengths above 150~G are preferentially concentrated in the bright, magnetically complex AR core, where PFSS is expected to be least reliable. However, repeating the analysis after excluding the AR core does not remove the elbow-like behaviour, indicating that the break is not solely caused by the core loops dominating the sample. We therefore regard the decrease in FIP bias at high mean field strength as a robust feature of this dataset.}

\rev{At present, however, we do not have a firm theoretical explanation for why the transition should occur specifically near 150~G. Recent MHD simulations of the ponderomotive force by Martinez-Sykora et al.~\cite{Martinez-Sykora2023ApJ...949..112M} suggest that in regions of strong magnetic flux density, where ion--neutral coupling is reduced, upward-propagating Alfv\'en waves can produce a downward ponderomotive force that depletes low-FIP elements. This provides a qualitative framework for understanding why FIP bias may decrease in strong-field regions, in agreement with our observed high-field trend. However, these models do not predict a clear universal threshold at 150~G. Therefore, we interpret this value primarily as an empirical transition identified in AR~11967 rather than as a theoretically established critical magnetic field strength. Determining whether a similar transition appears at comparable field strengths in other ARs will require a larger observational sample and more direct comparison with fractionation models.}

When comparing diagnostics without S involvement (Ca/Ar) to those including S (Si/S, Fe/S, and S/Ar), we find evidence that S does not consistently behave as a high-FIP element relative to Ar, with S likely depleted in regions of high magnetic field strength. The slope of S/Ar beyond 150~G supports our interpretation that S behaves more variably than a typical high-FIP element. Our results show for the first time that S, similar to other low-FIP elements, is more depleted in high magnetic field regions.

The ponderomotive force model predicts an enhanced S fractionation in slow solar wind,  which is associated with long open field magnetic field lines. However, we observe no clear correlation between S abundance and either loop length or open field lines. We note that for the 9 rasters studied here, the footpoints of open magnetic field associated with this AR were outside the EIS FOV, making it unclear whether this absence of correlation is a physical result or a limitation of our dataset. The lack of open-field regions with S depletion in our sample may therefore reflect observational coverage rather than the absence of S fractionation in solar wind source regions.

Our solar results also complement stellar composition studies. FIP bias has been shown to correlate with magnetic activity on both the Sun and other stars~\cite{Brooks2017Aug,Seli2022A&A...659A...3S}. Instead of the FIP effect (enhancement of low-FIP elements abundance), the inverse-FIP effect (depletion of low-FIP elements) is frequently observed in e.g., late-type binaries or M-dwarf stellar coronae~\cite{Testa2010SSRv..157...37T,Testa2015RSPTA.37340259T,Wood2018ApJ...862...66W, Seli2022A&A...659A...3S, 2025A&A...695A.165C}. At the same time, M-dwarf stars often exhibit higher surface magnetic field strengths ranging from $\sim$1kG--7kG~\cite{Reiners2007ApJ...656.1121R, Shulyak2017NatAs...1E.184S}. If we extrapolate our observed FIP bias trend above 150~G to the much stronger field strengths seen in M-dwarfs, we expect even more pronounced depletion of elements including Ca, Si and S towards the inverse FIP bias values (<1). This suggests that the fractionation process could be consistent for both the solar and stellar atmosphere, and solar studies can inform us about the stellar atmosphere.

Building on this, a future continuation of this study would be to establish a quantitative connection between the observed mean magnetic field strength dependence and existing theoretical fractionation models. In parallel, extending the dataset to include a large sample of ARs will also be important for testing whether the transition near 150~G is a general property of AR composition or a feature specific to AR~11967.

\section{Appendix}

\subsection{Identifying Diagnostic Line Pairs}

\begin{figure}[htbp]
    \centering
    \includegraphics[width=\linewidth]{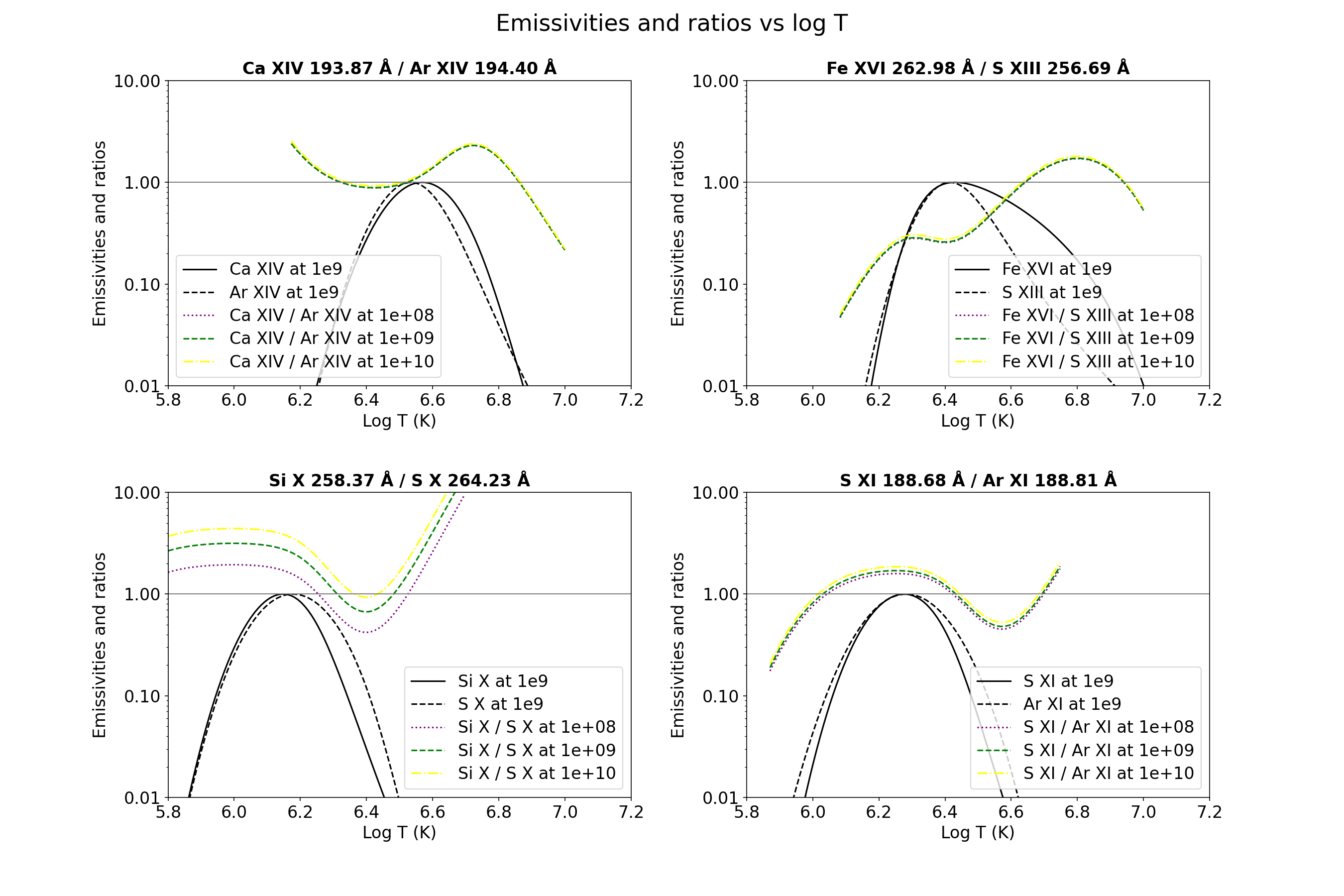}
    \caption{Emissivity and ratio curves for the four diagnostic line pairs used in this study: Si\,\textsc{x} 258.37~\AA\ / S\,\textsc{x} 264.23~\AA, S\,\textsc{xi} 188.68~\AA\ / Ar\,\textsc{xi} 188.81~\AA, Ca\,\textsc{xiv} 193.87~\AA\ / Ar\,\textsc{xiv} 194.40~\AA, and Fe\,\textsc{xvi} 262.98~\AA\ / S\,\textsc{xiii} 256.69~\AA. The solid curves show emissivity versus temperature, and the dashed curves show emissivity ratios at three electron densities. Pairs with overlapping emissivity peaks and stable ratio curves across density were selected for FIP bias analysis.}
    \label{fig:emissivities_ratio}
\end{figure}

Figure~\ref{fig:emissivities_ratio} shows the emissivity and ratio curves used to select the spectral line pairs. We selected line pairs based on overlapping emissivity peaks and stable ratio curves with low density sensitivity to ensure robust FIP bias diagnostics. Emissivity curves were calculated using the~\texttt{Chiantipy} atomic database.

\subsection{Intensity Maps}

\begin{figure}[htbp]
    \centering
    \includegraphics[width=\linewidth]{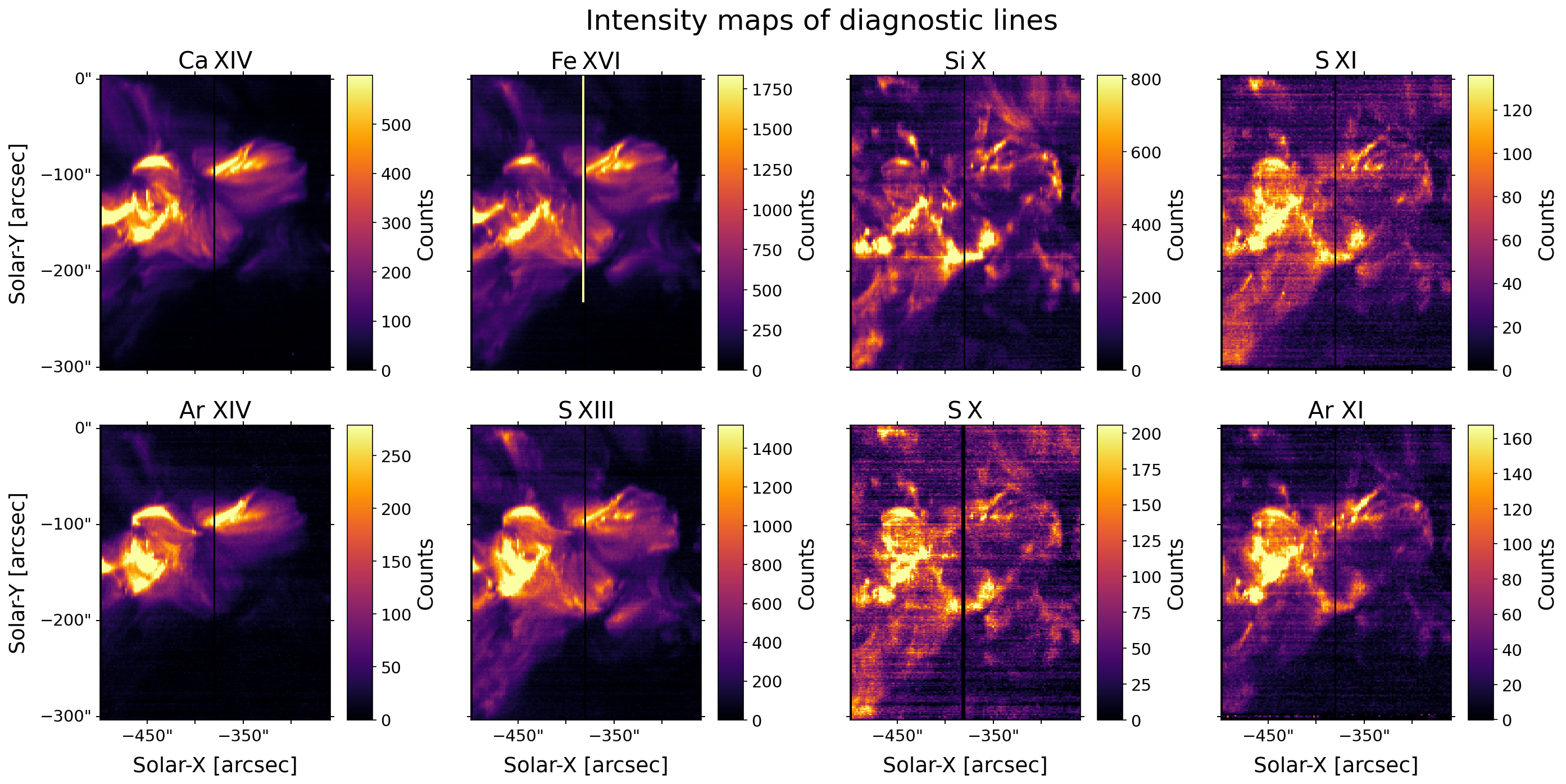}
    \caption{Intensity maps for the Ca~XIV, Fe~XVI, Si~X, S~XI, Ar~XIV, S~XIII, S~X, and Ar~XI diagnostic lines, derived from raw~\textit{Hinode}/EIS observations on 2014-02-01T10:50. These maps were used for emission structure analysis and for intensity-based filtering.}
    \label{fig:Intensity_Maps_2014_02_01_105035}
\end{figure}
\begin{figure}[htbp]
    \centering
    \includegraphics[width=\linewidth]{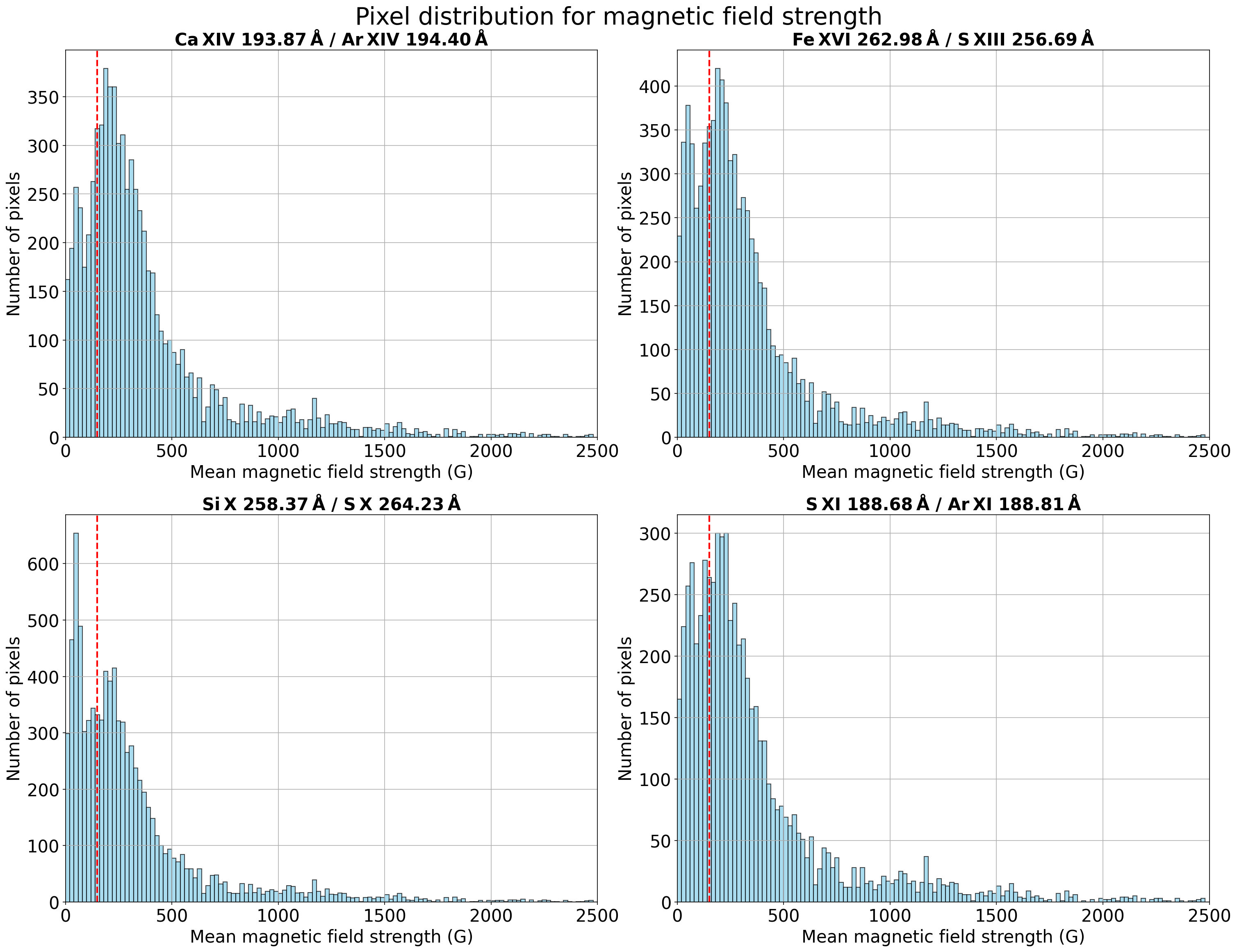}
    \caption{Distribution of valid closed field line pixels by mean magnetic field strength for all nine rasters combined, shown separately for each diagnostic line pair. The vertical line marks the 150~G threshold used for the two-slope fits in Figure~\ref{fig:Abundance_mean_magnetic}.}
    \label{fig:histogram}
\end{figure}
Figure~\ref{fig:Intensity_Maps_2014_02_01_105035} shows the intensity maps for the Ca~XIV, Fe~XVI, Si~X, S~XI, Ar~XIV, S~XIII, S~X, and Ar~XI diagnostic lines, derived from raw \textit{Hinode}/EIS observations on 2014-02-01T10:50:35. These maps were used for emission structure analysis and for intensity-based filtering. The Fe XII 195.12~\AA\ line was used as a stable reference for spatial alignment.

\subsection{Pixel Distribution for Magnetic Field Strength}
\begin{figure}[htbp]
    \centering
    \includegraphics[width=\linewidth]{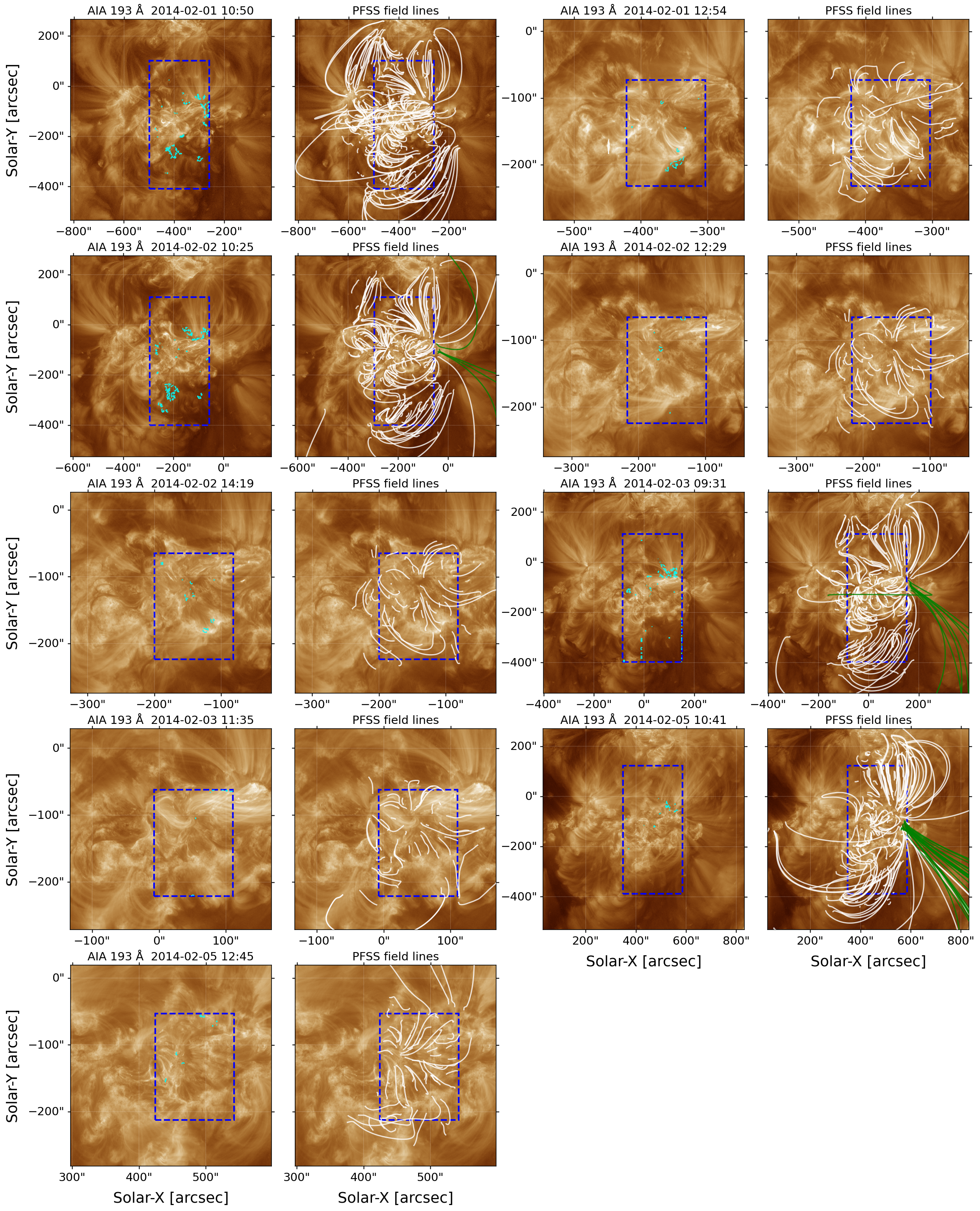}
    \caption{\rev{AIA 193~\AA\ context images for all nine EIS rasters. For each raster time (left to right): AIA 193~\AA\ image with Fe~XII 195.12 Doppler upflow contours (cyan; $\le -10\, \mathrm{km\,s^{-1}}$) and the EIS FOV (blue dashed box), and the corresponding AIA image with the PFSS field lines overlaid. Closed field lines are white, open field lines green.}}
    \label{fig:All_AIA_PFSS}
\end{figure}
Figure~\ref{fig:histogram} shows the pixel count distribution for closed field lines as a function of mean magnetic field strength for each diagnostic line pair. The majority of pixels lie below 500~G, with a relatively even distribution between the regions above and below 150~G threshold used in Figure~\ref{fig:Abundance_mean_magnetic}. However the Ca/Ar distribution is not as even and is noticeably weighted towards $\geq$~150~G.

\subsection{AIA 193~\AA\ with PFSS Overlays for All Rasters}

\rev{Figure~\ref{fig:All_AIA_PFSS} shows AIA 193~\AA\ context images and PFSS field line overlays for all nine EIS rasters. Across all observations, the PFSS extrapolation indicates that open field lines lay predominately outside the EIS FOV, while Fe~XII Doppler upflow contours within the FOV show little spatial overlap with the open field lines, consistent with the behaviour we see in Figure~\ref{fig:HMI_AIA_PFSS}.}

\subsection{FIP Maps with PFSS Overlays for All Rasters}

\begin{figure}[htbp]
    \centering
    \includegraphics[width=\linewidth]{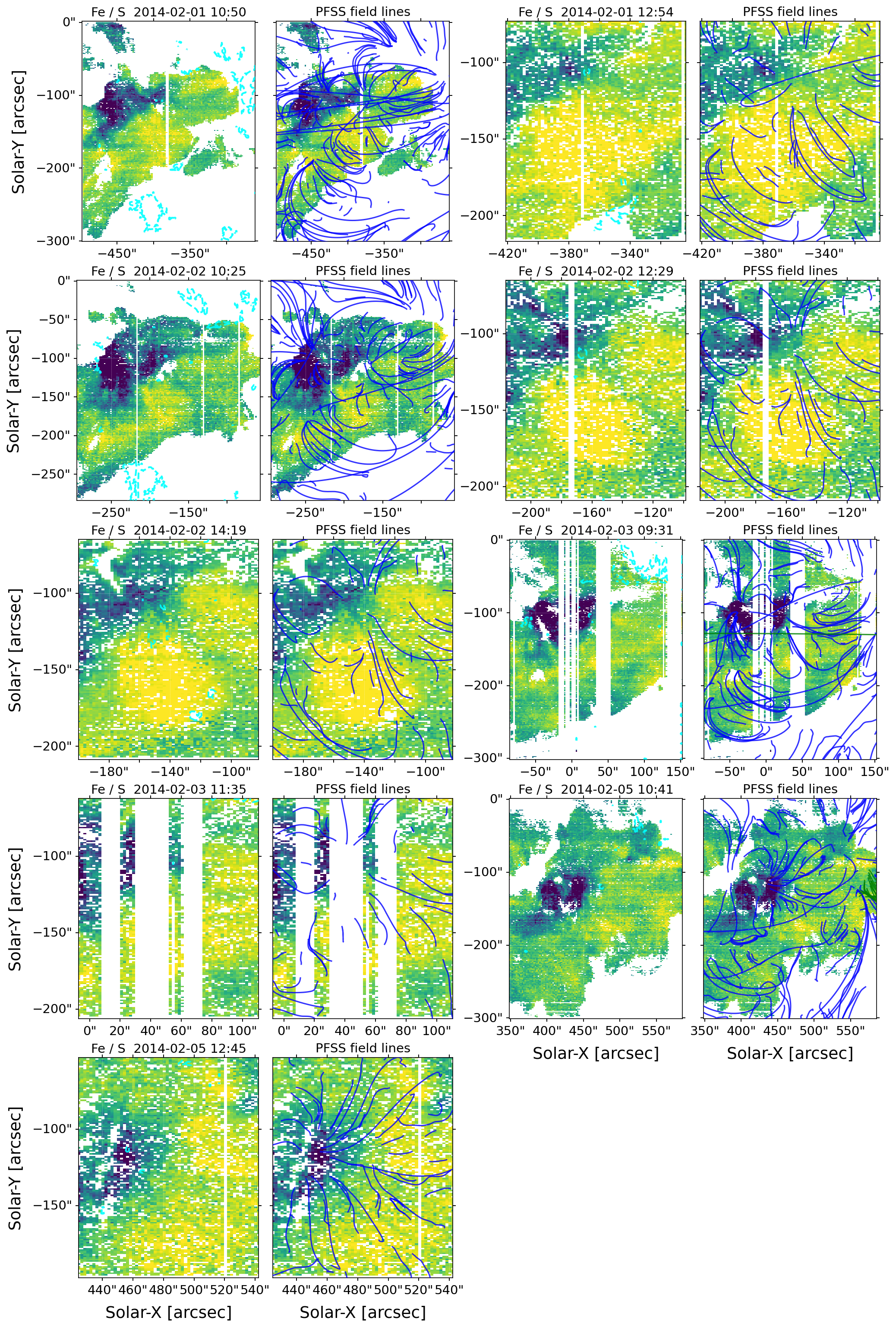}
    \caption{\rev{Fe/S FIP bias maps for all nine EIS rasters. For each raster time (left to right): the Fe/S map with Fe~XII 195.12 Doppler upflow contours (cyan; $\le -10\, \mathrm{km\,s^{-1}}$), and the corresponding map with the PFSS field lines overlaid. Closed field lines are blue, open field lines green.}}
    \label{fig:All_FIP_PFSS}
\end{figure}

\rev{Figure~\ref{fig:All_FIP_PFSS} shows Fe/S FIP bias maps and PFSS field line overlays for all nine EIS rasters. The overlays are visually crowded; motivating the statistical analysis presented in Figures~\ref{fig:Abundance_length} and~\ref{fig:Abundance_mean_magnetic}.}

\subsection{AIA 193~\AA\ with PFSS Overlays Above 150~G}

\begin{figure}[htbp]
    \centering
    \includegraphics[width=\linewidth]{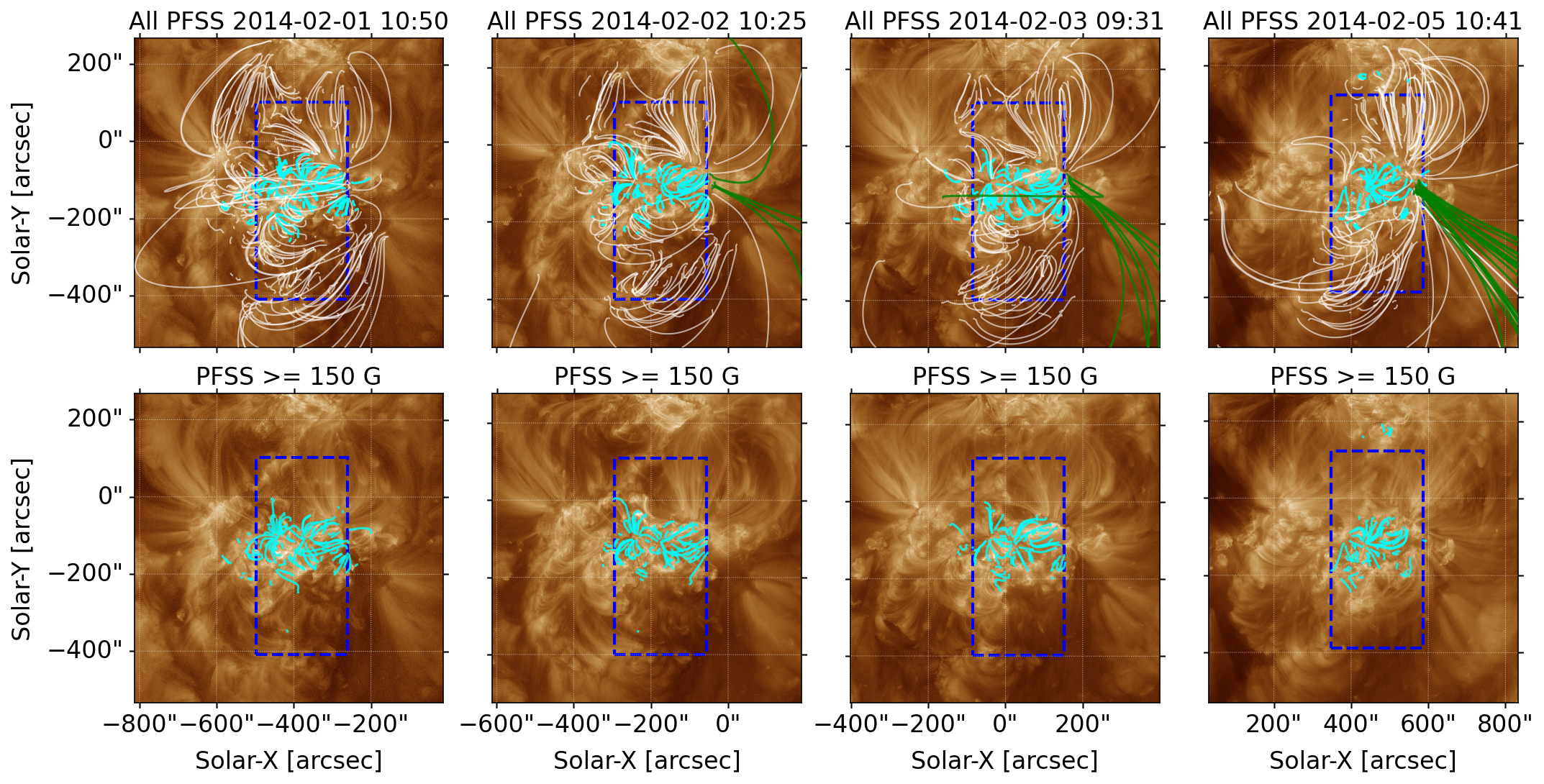}
    \caption{\rev{AIA 193~\AA\ context for the four large-FOV rasters used in the mean magnetic field analysis. The top row shows all traced PFSS field lines, with closed field lines of mean magnetic filed $\geq 150$~G highlighted in cyan. The bottom row shows only the closed PFSS field lines with mean magnetic field $\geq 150$~G. Open field lines are shown in green, other closed field lines in white, and the EIS FOV outlined in blue.}}
    \label{fig:AIA_PFSS_above150g_Appendix}
\end{figure}

\rev{Figure~\ref{fig:AIA_PFSS_above150g_Appendix} suggests that the strongest closed PFSS field lines are preferentially concentrated around the AR region core.}



\dataccess{The Hinode/EIS level-1 data used in this study are publicly
available from the Naval Research Laboratory (NRL) Hinode/EIS database
(\url{https://eis.nrl.navy.mil/}). Custom analysis scripts and
associated analysis materials are available from Zenodo as Orlovskij
D, To ASH, Long DM. 2026. Signs of Sulphur fractionation under high
magnetic field strength. Zenodo. Version v1.
doi:10.5281/zenodo.19946206.}

\ack{
\rev{We thank the referees for the insightful comments that improved this manuscript.} DO would like to thank David M. Long for organising his INTRA internship opportunity at Dublin City University and for establishing the collaboration with the European Space Agency. DO is also very grateful to Andy S.~H.~To for supervising the project and for his invaluable guidance and support throughout this work. ASHT acknowledges support through the European Space Agency (ESA) Research Fellowship Programme in Space Science.
Hinode is a Japanese mission developed and launched by ISAS/JAXA collaborating with NAOJ as a domestic partner and NASA and STFC (UK) as international partners. Scientific operation of the Hinode missions is conducted by the Hinode science team organised at ISAS/JAXA. This paper made use of several open source packages including astropy~\cite{AstropyCollaboration2022ApJ...935..167A}, sunpy~\cite{Mumford2020JOSS....5.1832M, SunPyCommunity2020ApJ...890...68S}, matplotlib~\cite{Hunter2007CSE.....9...90H}, numpy~\cite{Harris2020Natur.585..357H}, scipy~\cite{Virtanen2020NatMe..17..261V}, EISPAC~\cite{Weberg2023JOSS....8.4914W}, and \texttt{PFSSpy}~\cite{2020JOSS....5.2732S}
}

\bibliographystyle{RS}  
\bibliography{references}

\end{document}